\documentclass[11pt]{article}

\usepackage[left=1in, right=1in, top=1in, bottom=1in]{geometry}
\usepackage[group-separator={,}]{siunitx}
\usepackage{amsmath,amsfonts,amssymb, amsthm}
\usepackage{graphicx}
\usepackage{url}
\usepackage{mathtools}
\usepackage[textsize=tiny,textwidth=1cm,shadow]{todonotes}
\usepackage{thmtools}
\usepackage{enumitem}
\usepackage{pdflscape}
\usepackage{verbatim}
\usepackage{nccmath}
\usepackage[utf8]{inputenc}
\usepackage{array}
\definecolor{MyBlue}{rgb}{0.12, 0.12, 0.76}
\usepackage[colorlinks,allcolors=MyBlue]{hyperref}

\usepackage{algorithm}
\usepackage{algpseudocode}
\usepackage{algorithmicx}
\let\oldReturn\Return
\renewcommand{\Return}{\State\oldReturn}
\algtext*{EndWhile}
\algtext*{EndIf}
\algtext*{EndForAll}
\algtext*{EndFor}
\algtext*{EndFunction}

\usepackage{pifont}
\usepackage{pgf}
\usepackage{tikz}
\usetikzlibrary{shadows,arrows,decorations,decorations.shapes,backgrounds,shapes,snakes,automata,fit,petri,shapes.multipart,calc,positioning,shapes.geometric,graphs,graphs.standard}

\makeatletter
\newcommand{\thickhline}{%
    \noalign {\ifnum 0=`}\fi \hrule height 1.4pt
    \futurelet \reserved@a \@xhline
}
\newcolumntype{"}{@{\hskip\tabcolsep\vrule width 1.4pt\hskip\tabcolsep}}
\makeatother

\allowdisplaybreaks[1] 

\newtheorem{theorem}{Theorem}[section]

\usepackage{natbib}
\usepackage{tablefootnote} 

\newlist{exlist}{enumerate}{1}
\setlist[exlist]{label=(\alph*)}

\newcommand{\perf}{\textsc{Perf}}
\newcommand{\NN}{\mathbb{N}}
\newcommand{\eps}{\epsilon}
\newcommand{\RR}{\mathbb{R}}
\newcommand{\bfx}{\mathbf{x}}
\newcommand{\bfw}{\mathbf{w}}
\newcommand{\sse}{\subseteq}
\newcommand{\sgn}{\ensuremath{\mbox{sgn}}}
\newcommand{\n}[1]{
{\| {#1} \|}
}

\begin{document}

\title{Beyond the Worst-Case Analysis of Algorithms
  (Introduction)\thanks{Chapter~1 of the book {\em Beyond the
      Worst-Case Analysis of Algorithms}~\citep{bwca}.}}
\author{Tim Roughgarden\thanks{Department of Computer Science,
    Columbia University.  Supported in part by NSF award
    CCF-1813188 and ARO award W911NF1910294.  Email: \texttt{tim.roughgarden@gmail.com.}}}

\maketitle

\begin{abstract}
  One of the primary goals of the mathematical analysis of algorithms
  is to provide guidance about which algorithm is the ``best'' for
  solving a given computational problem.  Worst-case analysis
  summarizes the performance profile of an algorithm by its worst
  performance on any input of a given size, implicitly advocating for
  the algorithm with the best-possible worst-case performance.  Strong
  worst-case guarantees are the holy grail of algorithm design,
  providing an application-agnostic certification of an algorithm's
  robustly good performance.  However, for many fundamental problems
  and performance measures, such guarantees are impossible and a more
  nuanced analysis approach is called for.  This chapter surveys
  several alternatives to worst-case analysis that are discussed in
  detail later in the book.

\end{abstract}

\section{The Worst-Case Analysis of Algorithms}\label{s:intro}

\subsection{Comparing Incomparable Algorithms}

Comparing different algorithms is hard.  For almost any pair of
algorithms and measure of algorithm performance, 
each algorithm will perform better
than the other on some inputs.
For example, the MergeSort algorithm takes~$\Theta(n \log n)$ time to
sort length-$n$ arrays, whether the input is already sorted or not, while the
running time of the InsertionSort algorithm is~$\Theta(n)$ on
already-sorted arrays but $\Theta(n^2)$ in general.\footnote{A quick
  reminder about asymptotic notation in the analysis of algorithms:
  for nonnegative
  real-valued functions~$T(n)$ and~$f(n)$ defined on the natural
  numbers, we write
  $T(n) = O(f(n))$
if there are positive constants $c$ and $n_0$ such that $T(n)   \le c \cdot
f(n)$ for all $n \ge n_0$; $T(n) = \Omega(f(n))$ if there exist
positive $c$ and $n_0$ with $T(n) \ge c \cdot f(n)$ for all $n \ge
n_0$; and $T(n) = \Theta(f(n))$ if $T(n)$ is both $O(f(n))$ and $\Omega(f(n))$.}

The difficulty is not specific to running time analysis.  
In general, consider a computational problem $\Pi$ and a performance
measure $\perf$, with $\perf(A,z)$ quantifying the ``performance'' of
an algorithm~$A$ for~$\Pi$ on an input $z \in \Pi$.
For example,~$\Pi$ could be the Traveling Salesman Problem (TSP), $A$
could be a polynomial-time heuristic for the problem, and $\perf(A,z)$
could be the approximation ratio of~$A$---i.e., the ratio of the lengths of
$A$'s output tour and an optimal tour---on the TSP
instance~$z$.\footnote{In the Traveling Salesman Problem, the input is
  a complete undirected graph~$(V,E)$ with a nonnegative cost~$c(v,w)$
 for each edge $(v,w) \in E$, and the goal is to compute an ordering
$v_1,v_2,\ldots,v_n$ of the vertices~$V$ that minimizes the length
$\sum_{i=1}^n c(v_i,v_{i+1})$ of the corresponding tour
(with~$v_{n+1}$ interpreted as~$v_1$).}
Or~$\Pi$ could be the problem of testing primality, $A$
a randomized polynomial-time primality-testing algorithm, and
$\perf(A,z)$ the probability (over $A$'s internal randomness) that the
algorithm correctly decides if the positive integer~$z$ is prime.
In general, when two algorithms have incomparable performance, how can
we deem one of them ``better than'' the other?

{\em Worst-case analysis} is a specific modeling choice in the
analysis of algorithms, where the performance profile
$\{ \perf(A,z) \}_{z \in \Pi}$ of an algorithm is summarized by its
worst performance on any input of a given size (i.e.,
$\min_{z \,:\, |z|=n} \perf(A,z)$ or
$\max_{z \,:\, |z|=n} \perf(A,z)$, depending on the measure,
where~$|z|$ denotes the size of the input~$z$).  The ``better'' algorithm is
then the one with superior worst-case performance.  MergeSort, with
its worst-case asymptotic running time of $\Theta(n \log n)$ for
length-$n$ arrays, is better in this sense than InsertionSort, which
has a worst-case running time of $\Theta(n^2)$.

\subsection{Benefits of Worst-Case Analysis}

While crude, worst-case analysis can be tremendously useful and, for
several reasons, it has been
the dominant paradigm for algorithm analysis in theoretical computer
science.
\begin{itemize}

\item [1.]
A good worst-case guarantee is the best-case scenario for an
algorithm, certifying its general-purpose utility and absolving its
users from understanding which inputs are most relevant to their
applications.  
Thus worst-case analysis is particularly well suited for
``general-purpose'' algorithms that are expected to work well across a
range of application domains (like the default sorting routine of a
programming language).

\item [2.]
Worst-case analysis is often more analytically tractable to
carry out than its alternatives, such as average-case analysis with
respect to a probability distribution over inputs.

\item [3.] For a remarkable number of fundamental computational problems,
  there are algorithms with excellent worst-case performance
  guarantees.  For example, the lion's share of an undergraduate
  algorithms course comprises algorithms that run in linear or
  near-linear time in the worst case.\footnote{Worst-case analysis is also the dominant paradigm in
  complexity theory, where it has led to the development of
  $NP$-completeness and many other fundamental concepts.}

\end{itemize}

\subsection{Goals of the Analysis of Algorithms}

Before critiquing the worst-case analysis approach, it's worth
taking a step back to clarify 
why we want rigorous methods to reason about algorithm performance.
There are at least three possible goals:
\begin{itemize}

\item [1.] {\em Performance prediction.}
The first goal is to explain or predict the empirical performance of
algorithms.  
In some cases, 
the analyst acts as a natural scientist,
taking an observed phenomenon like ``the simplex method for linear
programming is fast'' as ground truth, and seeking a transparent
mathematical model that explains it.  
In others, the analyst plays the role of an engineer, seeking a theory
that gives accurate advice about whether or not an algorithm will
perform well in an application of interest.

\item [2.] {\em Identify optimal algorithms.}
The second goal is to rank different algorithms according to their
performance, and ideally to single out one algorithm as ``optimal.''
At the very least, given two algorithms $A$ and $B$ for the same
problem, a method for algorithmic analysis should offer an opinion
about which one is ``better.''

\item [3.] {\em Develop new algorithms.} The third goal is to provide a
  well-defined framework in which to brainstorm new algorithms.
Once a measure of algorithm performance has been declared, the
Pavlovian response of most computer scientists is to seek out new
algorithms that improve on the state-of-the-art with respect to
this measure.  The focusing effect catalyzed by such yardsticks should
not be underestimated.  

\end{itemize}
When proving or interpreting results in algorithm design and
analysis, it's important to be clear in one's mind about which of
these goals the work is trying to achieve.

What's the report card for worst-case analysis with respect to these
three goals?  
\begin{itemize}

\item [1.] Worst-case analysis gives an accurate performance prediction
  only for algorithms that exhibit little variation in performance
  across inputs of a given size.  This is the case for many of the
  greatest hits of algorithms covered in an undergraduate course,
  including the running times of near-linear-time algorithms and of
  many canonical dynamic programming algorithms.  For many more
  complex problems, however, the predictions of worst-case analysis are overly
  pessimistic (see Section~\ref{s:failures}).

\item [2.] For the
second goal, worst-case analysis earns a middling grade---it gives
good advice about which algorithm to use for some important problems
(like many of those in an undergraduate course) and bad advice for
others (see Section~\ref{s:failures}).  

\item [3.] Worst-case analysis has served as a tremendously useful
  brainstorming organizer. For over a half-century, researchers
  striving to optimize worst-case algorithm performance have been led
  to thousands of new algorithms, many of them practically useful.

\end{itemize}

\section{Famous Failures and the Need for Alternatives}\label{s:failures}

For many problems a bit beyond the scope of an undergraduate course,
the downside of worst-case analysis rears its ugly head.  
This section reviews four famous examples where worst-case analysis gives
misleading or useless advice about how to solve a problem.  These
examples motivate the alternatives to worst-case analysis that are
surveyed in Section~\ref{s:overview} and described in detail in 
later chapters of the book.

\subsection{The Simplex Method for Linear Programming}\label{ss:lp}

\begin{figure}
\begin{center}
\includegraphics[width=.35\textwidth]{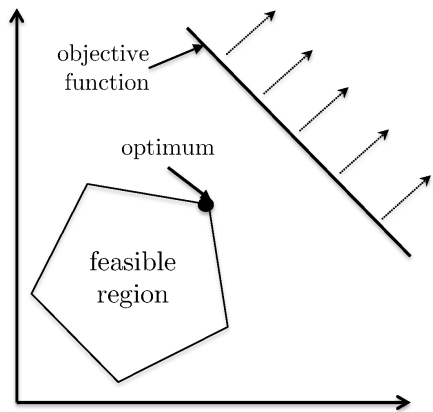}
\caption{A two-dimensional linear programming problem.}\label{f:lp}
\end{center}
\end{figure}

Perhaps the most famous failure of worst-case analysis concerns linear
programming, the problem of optimizing a linear function subject to
linear constraints (Figure~\ref{f:lp}).  
Dantzig proposed in the 1940s an
algorithm for solving linear programs called the {\em simplex method}.
The simplex method solves linear programs using greedy local search
on the vertices of the solution set boundary, and
variants of it remain in wide use to this day.  The enduring appeal of
the simplex method stems from its consistently superb performance in
practice.  Its running time typically scales modestly with the input
size, and it routinely solves linear programs with millions of
decision variables and constraints.  This robust empirical performance
suggested that the simplex method might well solve every linear
program in a polynomial amount of time.

\citet{KM72} 
showed by example that there are
contrived linear programs that force the simplex method to run in time
exponential in the number of decision variables (for all of the common
``pivot rules'' for choosing the next vertex).  This illustrates the
first potential pitfall of worst-case analysis: overly pessimistic
performance predictions that cannot be taken at face value.  The
running time of the simplex method is polynomial for all practical
purposes, despite the exponential prediction of worst-case analysis.

To add insult to injury, the first worst-case polynomial-time
algorithm for linear programming, the ellipsoid method,
is not competitive with the simplex method in
practice.\footnote{Interior-point methods,
developed five years later, lead to algorithms that both run in
  worst-case polynomial time and are competitive with the simplex
  method in practice.}
Taken at face value, worst-case analysis recommends the ellipsoid
method over the empirically superior simplex method.
One framework for narrowing the gap between these theoretical
predictions and empirical observations is {\em smoothed analysis},
the subject of Part~IV of this book; see Section~\ref{ss:sa} for an
overview.

\subsection{Clustering and $NP$-Hard Optimization Problems}\label{ss:clustering}

\begin{figure}
\centering
\includegraphics[width=.45\textwidth]{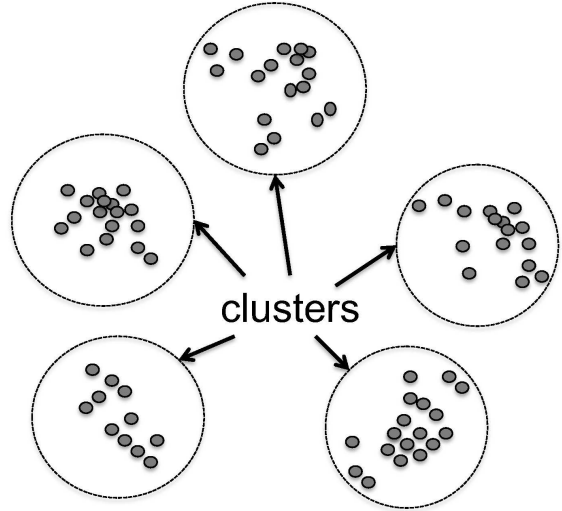}
\caption{A sensible clustering of a set of points.}\label{f:clustering}
\end{figure}

Clustering is a form of unsupervised learning (finding patterns in
unlabeled data), where the informal goal is to partition a set of
points into ``coherent groups'' (Figure~\ref{f:clustering}).  One
popular way to coax this goal into a well-defined computational
problem is to posit a numerical objective function over clusterings of
the point set, and then seek the clustering with the best objective
function value.  For example, the goal could be to choose $k$ cluster
centers to minimize the sum of the distances between points and their
nearest centers (the $k$-median objective) or the sum of the squared
such distances (the $k$-means objective).  Almost all natural
optimization problems that are defined over clusterings are
$NP$-hard.\footnote{Recall that a polynomial-time algorithm for an
  $NP$-hard problem would yield a polynomial-time algorithm for every
  problem in $NP$---for every problem with efficiently verifiable
  solutions.  Assuming the widely-believed $P \neq NP$ conjecture,
  every algorithm for an $NP$-hard problem either returns an
  incorrect answer for some inputs or runs in super-polynomial time
  for some inputs (or both).}

In practice, clustering is not viewed as a particularly difficult
problem.  Lightweight clustering algorithms, like Lloyd's algorithm
for $k$-means and its variants, regularly return the
intuitively ``correct'' clusterings of real-world point sets.  How can
we reconcile the worst-case intractability of clustering problems with
the empirical success of relatively simple algorithms?\footnote{More
  generally, optimization problems are more likely to be $NP$-hard
  than polynomial-time solvable.  In many cases, even computing an approximately optimal
  solution is an $NP$-hard problem.
  Whenever an efficient algorithm for such a problem
  performs better on real-world instances than (worst-case) complexity
  theory would suggest, there's an opportunity for a refined and more
  accurate theoretical analysis.}

One possible explanation is that {\em clustering is hard only when it
  doesn't matter}.
For example, if the difficult
instances of an $NP$-hard clustering problem look like a bunch of
random unstructured points, who cares?  The common use case for a
clustering algorithm is for points that represent images, or
documents, or proteins, or some other objects where a ``meaningful
clustering'' is likely to exist.  Could instances with a meaningful
clustering be easier than worst-case instances?  Part~III of this book
covers recent theoretical developments that support an affirmative
answer; see Section~\ref{ss:detmodels} for an overview.

\subsection{The Unreasonable Effectiveness of Machine Learning}\label{ss:ml}

The unreasonable effectiveness of modern machine learning algorithms
has thrown the gauntlet down to researchers in algorithm analysis, and
there is perhaps no other problem domain
that calls out as loudly for a ``beyond worst-case'' approach.

To illustrate some of the challenges, consider a canonical supervised
learning problem, where a learning algorithm is given a data set of
object-label pairs and the goal is to produce a classifier that
accurately predicts the label of as-yet-unseen objects (e.g., whether
or not an image contains a cat).  Over the past decade, aided by
massive data sets and computational power, neural networks have
achieved impressive levels of performance across a range of prediction
tasks.  Their empirical success flies in the face of conventional
wisdom in multiple ways.  First, there is a computational mystery:
Neural network training usually boils
down to fitting parameters (weights and biases) to minimize a
nonconvex loss function, for example to minimize the number of
classification errors the model makes on the training set.  In the
past such problems were written off as computationally intractable,
but first-order methods (i.e., variants of gradient descent) often
converge quickly to a local optimum or even to a global optimum.  Why?

Second, there is a statistical mystery: Modern neural networks are
typically over-parameterized, meaning that the number of parameters to
fit is considerably larger than the size of the training data set.
Over-parameterized models are vulnerable to large generalization error
(i.e., overfitting), since they can effectively memorize the training
data without learning anything that helps classify as-yet-unseen data
points.  Nevertheless, state-of-the-art neural networks generalize
shockingly well---why?  The answer likely hinges on special properties
of both real-world data sets and the optimization algorithms used for
neural network training (principally stochastic gradient descent).
Part~V of this book covers the state-of-the-art explanations of these
and other mysteries in the empirical performance of machine learning
algorithms.

The beyond worst-case viewpoint can also contribute to machine
learning by ``stress-testing'' the existing theory and providing a
road map for more robust guarantees.  While work in beyond worst-case
analysis makes strong assumptions relative to the norm in theoretical
computer science, these assumptions are usually weaker than the norm
in statistical machine learning.  Research in the latter field often
resembles average-case analysis, for example when data points are
modeled as independent and identically distributed samples from some
underlying structured distribution.
The semi-random models described in Parts~III and~IV of this book serve as
role models for blending adversarial and average-case modeling to
encourage the design of algorithms with robustly good performance.

\subsection{Analysis of Online Algorithms}\label{ss:paging}

{\em Online algorithms} are algorithms that must process their input
as it arrives over time.  For example, consider the online paging
problem, where there is a system with a small fast memory (the cache)
and a big slow memory.  Data is organized into blocks called {\em
  pages}, with up to $k$ different pages fitting in the cache at once.
A page request results in either a cache hit (if the page is already
in the cache) or a cache miss (if not).  On a cache miss, the
requested page must be brought into the cache.  If the cache is
already full, then some page in it must be evicted.  A cache
replacement policy is an algorithm for making these eviction
decisions.  Any systems textbook will recommend aspiring to the least
recently used (LRU) policy, which evicts the page whose most recent
reference is furthest in the past.  The same textbook will explain
why: Real-world page request sequences tend to exhibit locality of
reference, meaning that recently requested pages are likely to be
requested again soon.  The LRU policy uses the recent past as a
prediction for the near future.  Empirically, it typically suffers
fewer cache misses than competing policies like first-in first-out
(FIFO).

Worst-case analysis, straightforwardly applied, provides no useful
insights about the performance of different cache replacement
policies.  For every deterministic policy and cache size~$k$, there is
a pathological page request sequence that triggers a page fault rate
of 100\%, even though the optimal clairvoyant replacement policy
(known as B\'el\'ady's furthest-in-the-future algorithm) would have a
page fault rate of at most $1/k$ (Exercise~\ref{ex:paging}).  This
observation is troublesome both for its absurdly pessimistic
performance prediction and for its failure to differentiate between
competing replacement policies (like LRU vs.\ FIFO).  One solution,
described in Section~\ref{s:afg}, is to choose an appropriately
fine-grained parameterization of the input space and to assess and
compare algorithms using parameterized guarantees.

\subsection{The Cons of Worst-Case Analysis}

We should celebrate the fact that worst-case analysis works so well
for so many fundamental computational problems, while at the same time
recognizing that the cherrypicked
successes highlighted in undergraduate algorithms can paint a
potentially misleading picture
about the range of its practical relevance.
The preceding four examples highlight the chief weaknesses of the worst-case
analysis framework.
\begin{itemize}

\item [1.] {\em Overly pessimistic performance predictions.}  By design,
  worst-case analysis gives a pessimistic estimate of an algorithm's
empirical performance.  In the preceding four examples, 
the gap between the two is embarrassingly large.

\item [2.] {\em Can rank algorithms inaccurately.}  Overly pessimistic
  performance summaries can derail worst-case analysis from
  identifying the right algorithm to use in practice. In
the online paging problem, it cannot distinguish between the FIFO and LRU
policies; for linear programming, it implicitly suggests that the
ellipsoid method is superior to the simplex method.

\item [3.] {\em No data model.}  If worst-case analysis has an implicit
  model of data, then it's the ``Murphy's Law'' data model, where the
  instance to be solved is an adversarially selected function of the
  chosen algorithm.\footnote{Murphy's Law: If anything can go wrong,
    it will.}  Outside of security applications, this
  algorithm-dependent model of data is a rather paranoid and
  incoherent way to think about a computational problem.

  In many applications, the algorithm of choice is superior precisely
  because of properties of data in the application domain, like
  meaningful solutions in clustering problems or locality of reference
  in online paging.  Pure worst-case analysis provides no language for
  articulating such domain-specific properties of data.  In this
  sense, the strength of worst-case analysis is also its weakness.

\end{itemize}
These drawbacks show the importance of alternatives to worst-case
analysis, in the form of models that articulate properties of
``relevant'' inputs and algorithms that possess rigorous and
meaningful algorithmic guarantees for inputs with these properties.
Research in ``beyond worst-case analysis'' is a conversation between
models and algorithms, with each informing the development of the
other.  It has both a scientific
dimension, where the goal is to formulate transparent mathematical
models that explain empirically observed phenomena about algorithm
performance, and an engineering dimension, where the goals are to
provide accurate guidance about which algorithm to use for a problem
and to design new algorithms that perform particularly well on the
relevant inputs.

Concretely, what might a result that goes ``beyond worst-case
analysis'' look like?  The next section covers in detail an exemplary
result by \citet{AFG02} for the online paging problem introduced in
Section~\ref{ss:paging}.  The rest of the book offers dozens of
further examples.

\section{Example: Parameterized Bounds in Online Paging}\label{s:afg}

\subsection{Parameterizing by Locality of Reference}\label{ss:workingset}

Returning to the online paging example in Section~\ref{ss:paging},
perhaps we shouldn't be surprised that worst-case analysis fails to
advocate LRU over FIFO.  The empirical superiority of LRU is due to
the special structure in real-world page request sequences (locality
of reference), which is outside the language of pure worst-case
analysis.

The key idea for obtaining meaningful performance guarantees for and
comparisons between online paging algorithms is to parameterize page
request sequences according to how much locality of reference they
exhibit, and then prove parameterized worst-case guarantees.  Refining
worst-case analysis in this way leads to dramatically more informative
results.  Part~I of the book describes many other applications of such
fine-grained input parameterizations; see Section~\ref{ss:refine} for
an overview.

How should we measure locality in a page request sequence?
One tried and true method is the {\em working set} model,
which is parameterized by a function $f$
from the positive integers $\NN$ to $\NN$ that describes how many
different page requests are possible in a window of a given length.
Formally, we say that a page sequence~$\sigma$ {\em conforms to $f$}
if for every positive integer $n$ and every set of $n$ consecutive
page requests in $\sigma$, there are requests for at most $f(n)$ 
  distinct pages.  
For example, the identity function $f(n)=n$ imposes no restrictions on
the page request sequence.  A sequence can only conform to a sublinear
function like $f(n)=\lceil \sqrt{n} \rceil$ or $f(n) = \lceil 1 +
\log_2 n \rceil$ if it exhibits
locality of reference.\footnote{The notation $\lceil x \rceil$ means
  the number $x$, rounded up to the nearest integer.}
We can assume without loss of generality that 
$f(1)=1$, $f(2)=2$, and $f(n+1) \in \{ f(n), f(n)+1\}$ for all $n$
(Exercise~\ref{ex:concave}).

We adopt as our performance measure $\perf(A,z)$ the fault rate of an
online algorithm~$A$ on the page request sequence~$z$---the fraction
of requests in~$z$ on which~$A$ suffers a page fault.  We next state a
performance guarantee for the fault rate of the LRU policy with a
cache size of~$k$ that is parameterized by a number
$\alpha_f(k) \in [0,1]$.  The parameter $\alpha_f(k)$ is defined below
in~\eqref{eq:alpha}; intuitively, it will be close to~0 for
slow-growing functions~$f$ (i.e., functions that impose strong
locality of reference) and close to~1 for functions~$f$ that grow
quickly (e.g., near-linearly).  This performance guarantee
requires that the function~$f$ is {\em approximately concave} in the
sense that the number~$m_y$ of inputs with value~$y$ under~$f$ (that
is, $|f^{-1}(y)|$) is nondecreasing in~$y$ (Figure~\ref{f:concave}).

\begin{figure}
\begin{center}
\begin{tabular}{c|c|c|c|c|c|c|c|c|c}
$f(n)$ & 1 & 2 & 3 & 3 & 4 & 4 & 4 & 5 & $\cdots$ \\ \hline
   $n$ & 1 & 2 & 3 & 4 & 5 & 6 & 7 & 8 & $\cdots$ \\
\end{tabular}
\caption{An approximately concave function, with $m_1 = 1$, $m_2 = 1$,
  $m_3 = 2$, $m_4 = 3, \ldots$}\label{f:concave}
\end{center}
\end{figure}

\begin{theorem}[\citet{AFG02}]\label{t:afg}

With $\alpha_f(k)$ defined as in~\eqref{eq:alpha} below:
\begin{itemize}

\item [(a)] For every approximately concave function $f$, 
  cache size $k \ge 2$,
and deterministic cache replacement policy, there are arbitrarily long
page request sequences conforming to~$f$ for which the policy's page
fault rate is at least $\alpha_f(k)$.

\item [(b)] For every approximately concave function $f$, 
  cache size $k \ge 2$, and page request sequence that conforms to $f$,
the page fault rate of the LRU policy is at most $\alpha_f(k)$ plus
an additive term that goes to~0 with the sequence length.

\item [(c)] There exists a choice of an approximately concave function
  $f$, a cache size $k \ge 2$, and an arbitrarily long page request
  sequence that conforms to~$f$, such that the
page fault rate of the FIFO policy is bounded away from $\alpha_f(k)$.

\end{itemize}
\end{theorem}
Parts~(a) and~(b) prove the worst-case optimality of the LRU policy
in a strong and fine-grained sense, $f$-by-$f$ and $k$-by-$k$.
Part~(c) differentiates LRU from FIFO, as the latter is suboptimal for
some (in fact, many) choices of $f$ and $k$.

The guarantees in Theorem~\ref{t:afg} are so
good that they are meaningful even when taken at face value---for
strongly sublinear~$f$'s, $\alpha_f(k)$ goes to~0 reasonably quickly
with~$k$. 
The precise definition of $\alpha_f(k)$ for $k \ge 2$ is
\begin{equation}\label{eq:alpha}
\alpha_f(k) = \frac{k-1}{f^{-1}(k+1)-2},
\end{equation}
where we abuse notation and interpret $f^{-1}(y)$ as the smallest
value of $x$ such that $f(x) = y$.
That is, $f^{-1}(y)$ denotes the smallest window length in which page
requests for $y$ distinct pages might appear.
As expected, for the function $f(n)=n$ we have $\alpha_f(k) = 1$ for
all~$k$.  (With no restriction on the input sequence, an adversary can
force a 100\% fault rate.)
If $f(n) = \lceil \sqrt{n} \rceil$, however, then $\alpha_f(k)$ scales 
with $1/\sqrt{k}$.
Thus with a cache size of 10,000, the page fault
rate is always at most 1\%.
If $f(n) = \lceil 1 + \log_2 n \rceil$, then $\alpha_f(k)$ goes to~0
even faster with~$k$, roughly as $k/2^k$.

\subsection{Proof of Theorem~\ref{t:afg}}

This section proves the first two parts of Theorem~\ref{t:afg};
part~(c) is left as Exercise~\ref{exer:afg}.

\vspace{-.1in}
\paragraph{Part~(a).}
To prove the lower bound in part~(a), fix an approximately concave
function $f$ and a cache size~$k \ge 2$.  
Fix a deterministic cache replacement policy $A$.  

We construct a page sequence $\sigma$ that uses only $k+1$ distinct
pages, so at any given time step there is exactly one page missing
from the algorithm's cache.  (Assume that the algorithm begins with
the first~$k$ pages in its cache.)  
The sequence comprises $k-1$ blocks,
where the $j$th block consists of $m_{j+1}$
consecutive requests for the same page $p_j$, where $p_j$ is the
unique page missing from the algorithm $A$'s cache at the start of the
block.  (Recall that $m_y$ is the number of values of~$x$ such that
$f(x)=y$.)  
This sequence conforms to~$f$
(Exercise~\ref{ex:conform}).

By the choice of the $p_j$'s, $A$ incurs a page fault on
the first request of a block, and not on any of the other (duplicate)
requests of that block.  Thus, algorithm $A$ suffers exactly $k-1$
page faults.  

The length of the page request sequence is $m_2 + m_3 + \cdots + m_k$.
Because $m_1 = 1$, this sum equals $(\sum_{j=1}^k m_j) - 1$ which,
using the definition of the $m_j$'s, equals
$(f^{-1}(k+1)-1)-1 = f^{-1}(k+1)-2$.  The algorithm's page fault rate
on this sequence matches the definition~\eqref{eq:alpha} of
$\alpha_f(k)$, as required.  More generally, repeating the
construction over and over again produces arbitrarily long page
request sequences for which the algorithm has page fault
rate~$\alpha_f(k)$.

\begin{figure}
\begin{center}
\includegraphics[width=.85\textwidth]{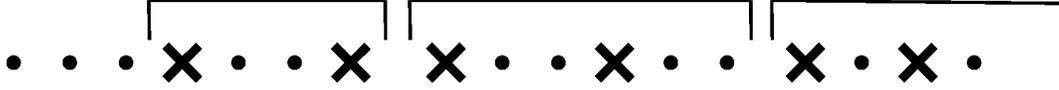}
\caption{Blocks of $k-1$ faults, for $k=3$.}\label{f:ub}
\end{center}
\end{figure}

\vspace{-.1in}
\paragraph{Part~(b).}
To prove a matching upper bound for the LRU policy, fix an
approximately concave function $f$, a cache size $k \ge 2$, and a
sequence $\sigma$ that conforms to $f$.  Our fault rate target
$\alpha_f(k)$ is a major clue to the proof (recall~\eqref{eq:alpha}):
we should be looking to partition the sequence $\sigma$ into blocks of
length at least $f^{-1}(k+1)-2$ such that each block has at most $k-1$
faults.
So consider groups of $k-1$ consecutive faults of the LRU policy on
$\sigma$.  Each such group defines a {\em block}, beginning with the
first fault of the group, and ending with the page request that
immediately precedes the beginning of the next group of faults (see
Figure~\ref{f:ub}).

\vspace{.1in}
\noindent
{\bf Claim: } Consider a block other than the first or last.  Consider
the page requests in this block, together with the requests
immediately before and after this block.  These requests 
are for at least $k+1$ distinct pages.

\vspace{.1in}

The claim immediately implies that every block contains at least
$f^{-1}(k+1)-2$ requests.  Because there are $k-1$ faults per block, this
shows that the page fault rate is at most $\alpha_f(k)$ (ignoring the
vanishing additive error due to the first and last blocks), 
proving Theorem~\ref{t:afg}(b).

We proceed to the proof of the claim.  Note that, in light of
Theorem~\ref{t:afg}(c), it is essential that the proof uses
properties of the LRU policy not shared by FIFO.  Fix 
a block other than the first or last, and let $p$ be the page requested
immediately prior to this block.  This request could have
been a page fault, or not (cf., Figure~\ref{f:ub}).  In any case,
$p$ is in the cache when this block begins.
Consider the $k-1$ faults contained in the block, together with the
$k$th fault that occurs immediately after the block.
We consider three cases.

First, if the $k$ faults occurred on distinct pages that are all
different from $p$, we have identified our $k+1$
distinct requests ($p$ and the $k$ faults).  For the second case,
suppose that two of the $k$ faults were for the same page $q \neq p$.
How could this have happened?
The page~$q$ was brought into the cache after the first fault on~$q$,
and was not evicted until there were $k$ requests for distinct pages
other than $q$ after this page fault.  This gives $k+1$ distinct page
requests ($q$ and the $k$ other distinct requests between the two
faults on $q$).  Finally, suppose that one of these~$k$ faults was on
the page $p$.  Because $p$ was requested just before the first of
these faults, the LRU algorithm, subsequent to this request and prior
to evicting $p$, must have received requests for $k$ distinct pages
other than $p$.  These requests, together with that for $p$, give the
desired $k+1$ distinct page requests.\footnote{The first two arguments
  apply also to the FIFO policy, but the third does not.  Suppose
  $p$ was already in the cache when it was requested just prior to the
  block.  Under FIFO, this request does not ``reset $p$'s clock''; if
  it was originally brought into the cache long ago, FIFO might well
  evict $p$ on the block's very first fault.}

\subsection{Discussion}

Theorem~\ref{t:afg} is an example of a ``parameterized analysis'' of an
algorithm, where the performance guarantee is expressed as a function
of parameters of the input other than its size.
A parameter like~$\alpha_f(k)$ measures the ``easiness'' of an input,
much like matrix condition numbers in linear algebra.  We will see
many more examples of parameterized analyses later in the book.

There are several reasons to aspire toward parameterized performance
guarantees.
\begin{itemize}

\item [1.] A parameterized guarantee is a mathematically stronger
  statement, containing strictly more information about an algorithm's
  performance than a worst-case guarantee parameterized solely by the
  input size.

\item [2.] A parameterized analysis can explain why an algorithm has good
  ``real-world'' performance even when its worst-case performance is
  poor. The approach is to first show that the algorithm performs well
  for ``easy'' values of the parameter (e.g., for $f$ and $k$ such
  that $\alpha_f(k)$ is close to~0), and then make a case that
  ``real-world'' instances are ``easy'' in this sense (e.g., have
  enough locality of reference to conform to a function~$f$ with a
  small value of~$\alpha_f(k)$).  The latter argument can be made
  empirically (e.g., by computing the parameter on representative
  benchmarks) or mathematically (e.g., by positing a generative model
  and proving that it typically generates “easy” inputs).  Results in
  smoothed analysis (see Section~\ref{ss:sa} and Part~IV) typically
  follow this two-step approach.

\item [3.] A parameterized performance guarantee suggests when---for which
  inputs, and which application domains---a given algorithm should be
  used.  (Namely, on the inputs where
the performance of the algorithm is good!) 
Such advice is useful to 
someone who has no time or interest in developing their own 
algorithm from scratch, and merely wishes to be an
educated client of existing algorithms.\footnote{For a familiar example, 
  parameterizing the running time of graph algorithms by both
  the number of vertices and the number of edges provides guidance
  about which algorithms should be used for sparse graphs and which 
ones for dense graphs.}

\item [4.] Fine-grained performance characterizations can differentiate
  algorithms when worst-case analysis cannot (as with LRU vs.\ FIFO).

\item [5.] Formulating a good parameter often forces the analyst to
  articulate a form of structure in data, like the ``amount of
  locality'' in a page request sequence.  Ideas for new algorithms
  that explicitly exploit such structure often follow soon
  thereafter.\footnote{The parameter~$\alpha_f(k)$ showed up only in
    our {\em analysis} of the LRU policy; in other applications,
    the chosen parameter also guides the {\em design} of algorithms
    for the problem.}

\end{itemize}

Useful parameters come in several
flavors.  The parameter~$\alpha_f(k)$ in Theorem~\ref{t:afg} is
derived directly from the input to the problem, and later chapters 
contain many more examples
of such input-based parameters.
It is also common to parameterize algorithm performance by properties
of an optimal solution.  In parameterized algorithms (Chapter~2), the
most well-studied such parameter is the size of an optimal solution.
Another solution-based parameterization, popular in machine learning
applications, is by the ``margin,'' meaning the extent to which the
optimal solution is particularly ``pronounced''; see
Exercise~\ref{ex:perceptron} for the canonical example of the
analysis of the perceptron algorithm.

``Input size'' is well defined for every computational problem, and this
is one of the reasons why performance guarantees parameterized by
input size are so ubiquitous.  By contrast, the
parameter~$\alpha_f(k)$ used in Theorem~\ref{t:afg} is specifically
tailored to the online paging problem; in exchange, the performance
guarantee is unusually accurate and meaningful.
Alas, there are no silver bullets in parameterized analysis, or in
algorithm analysis more generally, and the most enlightening analysis
approach is often problem-specific.  Worst-case analysis can inform
the choice of an appropriate analysis framework for a problem by
highlighting the problem's most difficult (and often unrealistic)
instances.

\section{Overview of Book}\label{s:overview}

This book has six parts, four on ``core theory'' and two on
``applications.''  Each of the following sections summarizes the
chapters in one of the parts.

\subsection{Refinements of Worst-Case Analysis}\label{ss:refine}

Part~I of the book hews closest to traditional worst-case analysis.
No assumptions are imposed on the input; as in worst-case analysis,
there is no commitment to a ``model of data.''  The innovative ideas
in these chapters concern novel and problem-specific ways of
expressing algorithm performance.  Our online paging example
(Section~\ref{s:afg}) falls squarely in this domain.

Chapter~2, by Fomin, Lokshtanov, Saurabh and Zehavi, provides an
overview of the relatively mature field of {\em parameterized algorithms}.
The goal here is to understand how the running time of algorithms and the
complexity of computational problems depend on parameters other than
the input size.  For example, for which $NP$-hard problems $\Pi$ and
parameters~$k$ is $\Pi$ ``fixed-parameter tractable'' with respect
to~$k$, meaning solvable in time $f(k) \cdot n^{O(1)}$ for some
function~$f$ that is independent of the input size~$n$?   The field
has developed a number of powerful approaches to designing
fixed-parameter tractable algorithms, as well as lower bound
techniques for ruling out the existence of such algorithms
(under appropriate complexity assumptions).

Chapter~3, by Barbay, searches for {\em instance-optimal}
algorithms---algorithms that for every input perform better than every
other algorithm (up to a constant factor).  Such an input-by-input
guarantee is essentially the strongest notion of optimality one could
hope for.  Remarkably, there are several fundamental problems, for
example in low-dimensional computational geometry, that admit an
instance-optimal algorithm.  Proofs of instance optimality involve
input-by-input matching upper and lower bounds, and this typically
requires a very fine-grained parameterization of the input space.

Chapter~4, by Roughgarden, concerns {\em resource augmentation}.
This concept makes sense for problems that have a natural notion of
a ``resource,'' with the performance of an algorithm improving as it
is given more resources.  Examples include the size of a cache (with
larger caches leading to fewer faults), the capacity of a network
(with higher-capacity networks leading to less congestion), and the
speed of a processor (with faster processors leading to earlier job 
completion times).  A resource augmentation guarantee then states that
the performance of an algorithm of interest is always close to
that achieved by an all-powerful algorithm that is handicapped by
slightly less resources.

\subsection{Deterministic Models of Data}\label{ss:detmodels}

Part~II of the book proposes deterministic models of data for several
$NP$-hard clustering and sparse recovery problems, which effectively
posit conditions that are conceivably satisfied by ``real-world''
inputs.  This work fits into the long-standing tradition of
identifying ``islands of tractability,'' meaning polynomial-time
solvable special cases of $NP$-hard problems.  20th-century research
on tractable special cases focused primarily on syntactic and
easily-checked restrictions (e.g., graph planarity or Horn
satisfiability).  The chapters in Part~II and some of the related
application chapters consider conditions that are not necessarily
easy to check, but for which there is a plausible narrative about why
``real-world instances'' might satisfy them, at least approximately.

Chapter~5, by Makarychev and Makarychev, studies 
{\em perturbation stability} in several different computational
problems.  A perturbation-stable instance satisfies a property that is
effectively a uniqueness condition on steroids, stating that the
optimal solution remains invariant to sufficiently small perturbations
of the numbers in the input.  The larger the perturbations that are
tolerated, the stronger the condition on the instance and the easier
the computational problem.  Many problems have ``stability
thresholds,'' an allowable perturbation size at which the complexity
of the problem switches suddenly from $NP$-hard to polynomial-time
solvable.  To the extent that we're comfortable identifying
``instances with a meaningful clustering'' with perturbation-stable
instances, the positive results in this chapter give a precise sense
in which clustering is hard only when it doesn't matter (cf.,
Section~\ref{ss:clustering}).  As a bonus, many of these positive
results are achieved by algorithms that resemble popular approaches in
practice, like single-linkage clustering and local search.

Chapter~6, by Blum, proposes an alternative condition called {\em
  approximation stability}, stating that every solution with a
near-optimal objective function value closely resembles the optimal
solution.  
That is, any solution that is structurally different from
the optimal solution has significantly worse objective function value.
This condition is particularly appropriate for problems
like clustering, where the objective function is only means to an end
and the real goal is to recover some type of ``ground-truth'' clustering.
This chapter demonstrates that many $NP$-hard problems become provably
easier for approximation-stable instances.

Chapter~7, by Price, provides a glimpse of the vast literature on {\em
  sparse recovery}, where the goal is to reverse engineer a ``sparse''
object from a small number of clues about it.  This area is more
strongly associated with applied mathematics than with theoretical
computer science and algorithms, but there are compelling parallels
between it and the topics of the preceding two chapters.  For example,
consider the canonical problem in compressive sensing, where the goal
is to recover an unknown sparse signal~$z$ (a vector of length~$n$)
from a small number~$m$ of linear measurements of it.  If $z$ can be
arbitrary, then the problem is hopeless unless $m=n$.  But many
real-world signals have most of their mass concentrated on $k$
coordinates for small $k$ (and an appropriate basis), and the results
surveyed in this chapter show that, for such ``natural'' signals,
the problem can be solved efficiently even when~$m$ is only modestly
bigger than $k$ (and much smaller than $n$).

\subsection{Semi-Random Models}

Part~III of the book is about {\em semi-random models}---hybrids of
worst- and average-case analysis in which nature and an adversary
collaborate to produce an instance of a problem.  For many problems,
such hybrid frameworks are a ``sweet spot'' for algorithm
analysis, with the worst-case dimension encouraging the design of
robustly good 
algorithms and the average-case dimension allowing for strong provable
guarantees.

Chapter~8, by Roughgarden, sets the stage with a review of pure
average-case or {\em distributional analysis}, along with some of its
killer applications and biggest weaknesses.  Work in this area adopts a
specific probability distribution over the inputs of a problem, and
analyzes the expectation (or some other statistic) of the performance
of an algorithm with respect to this distribution.  One use of
distributional analysis is to show that a general-purpose algorithm has
good
performance on non-pathological inputs
(e.g., deterministic QuickSort on randomly ordered arrays).
One key drawback of distributional
analysis is that it can encourage the design of algorithms that are brittle
and overly tailored to the assumed input distribution.  The
semi-random models of the subsequent chapters are designed to
ameliorate this issue.

Chapter~9, by Feige, introduces several {\em planted models} and their
semi-random counterparts.  For example, in the planted clique problem,
a clique of size~$k$ is planted in an otherwise uniformly random
graph.  How large does~$k$ need to be, as a function of the number of
vertices, before the planted clique can be recovered in polynomial
time (with high probability)?  In a semi-random version of a planted
model, an adversary can modify the random input in a restricted way.
For example, in the clique problem, an adversary might be allowed to
remove edges not in the clique; such changes intuitively make the
planted clique only ``more obviously optimal,'' but nevertheless can
foil overly simplistic algorithms.  One rule of thumb that emerges
from this line of work, and also recurs in the next chapter, is that
spectral algorithms tend to work well for planted models but the
heavier machinery of semidefinite programming seems required for their
semi-random counterparts.  
This chapter also investigates random and semi-random models for Boolean
formulas, including refutation algorithms that certify that a
given input formula is not satisfiable.

Chapter~10, by Moitra, drills down on a specific and
extensively-studied planted model, the {\em stochastic block model}.
The vertices of a graph are partitioned into groups,
and each potential edge of the graph is present independently with a
probability that depends only on the groups that contain its
endpoints.  The algorithmic goal is to recover the groups from the
(unlabeled) graph.  One important special case is the planted
bisection problem, where the vertices are split into two equal-size
sets~$A$ and~$B$ and each edge is present independently with probability
$p$ (if both endpoints are in the same group) or $q < p$ (otherwise).
How big does the gap $p-q$ need to be before the planted bisection
$(A,B)$ can be recovered, either statistically (i.e., with unbounded
computational power) or with a polynomial-time algorithm?  When $p$
and $q$ are sufficiently small, the relevant goal becomes partial
recovery, meaning a proposed classification of the vertices with
accuracy better than random guessing.  In the semi-random version of
the model, an adversary can remove edges crossing the bisection and
add edges internal to each of the groups.  For partial recovery, this
semi-random version is provably more difficult than the original
model.

Chapter~11, by Gupta and Singla, describes results for a number of online
algorithms in {\em random-order models}.
These are semi-random models in which an adversary decides on 
an input, and nature then presents this input to an online algorithm, one
piece at a time and in random order.  The canonical example here is
the secretary problem, where an arbitrary finite set of numbers is
presented to an algorithm in random order, and the goal is to design a
stopping rule with the maximum-possible probability of stopping on the
largest number of the sequence.  Analogous random-order models have
proved useful for overcoming worst-case lower bounds for the online
versions of a number of combinatorial optimization 
problems, including bin packing, facility location, and network design.

Chapter 12, by Seshadhri, is a survey of the field of {\em
  self-improving algorithms}.  The goal here is to design an algorithm
that, when presented with a sequence of independent samples drawn from
an unknown input distribution, quickly converges to the optimal
algorithm for that distribution.  For example, for many distributions
over length-$n$ arrays, there are sorting algorithms that make less
than $\Theta(n \log n)$ comparisons on average.  Could there be a
``master algorithm'' that replicates the performance of a
distribution-optimal sorter from only a limited number of samples from
the distribution?  This chapter gives a positive answer under the
assumption that array entries are drawn independently (from possibly
different distributions), along with analogous positive results for
several fundamental problems in low-dimensional computational
geometry.

\subsection{Smoothed Analysis}\label{ss:sa}

Part~IV of the book focuses on the semi-random models studied
in {\em smoothed analysis}.  In smoothed analysis, an adversary chooses an
arbitrary input, and this input is then perturbed slightly by
nature.  The
performance of an algorithm is then assessed by its worst-case
expected performance, where the worst case is over the adversary's
input choice and the expectation is over the random perturbation.
This analysis framework can be applied to any problem where ``small
random perturbations'' make sense, including most problems with real-valued
inputs.  It can be applied to any measure of algorithm performance,
but has proven most effective for running time analyses of algorithms
that seem to run in super-polynomial time only on highly contrived
inputs (like the simplex method).  As with other semi-random models,
smoothed analysis has the benefit of potentially escaping worst-case
inputs, especially if they are ``isolated'' in the input space,
while avoiding overfitting a solution to a specific distributional
assumption.  There is also a plausible narrative about why
``real-world'' inputs are captured by this framework: Whatever problem
you'd like to solve, there are inevitable inaccuracies in its
formulation from measurement errors, uncertainty, and so on.

Chapter~13, by Manthey, details several applications of smoothed
analysis to the {\em analysis of local search algorithms} for
combinatorial optimization problems.  For example, the 2-opt heuristic
for the Traveling Salesman Problem is a local search algorithm that
begins with an arbitrary tour and repeatedly improves the current
solution using local moves that swap one pair of edges for another.
In practice, local search algorithms like the 2-opt heuristic almost
always converge to a locally optimal solution in a small number of
steps.  Delicate constructions show that the 2-opt heuristic, and many
other local search algorithms, require an exponential number of steps
to converge in the worst case.  The results in this chapter use
smoothed analysis to narrow the gap between worst-case analysis and
empirically observed performance, establishing that many local search
algorithms (including the 2-opt heuristic) have polynomial smoothed
complexity.

Chapter~14, by Dadush and Huiberts, surveys the first and most famous
killer application of smoothed analysis, the Spielman-Teng analysis of
the {\em running time of the simplex method} for linear programming.
As discussed in Section~\ref{ss:lp}, the running time of the simplex
method is exponential in the worst case but almost always polynomial in
practice.  This chapter develops intuition for and outlines a proof of
the fact that the simplex method, implemented with the shadow vertex pivot
rule, has polynomial smoothed complexity with respect to small
Gaussian perturbations of the entries of the constraint matrix.
The chapter also shows how to interpret the successive
shortest-path algorithm for the minimum-cost maximum-flow problem as an
instantiation of this version of the simplex method.

Chapter~15, by R\"oglin, presents a third application of smoothed
analysis, to the size of {\em Pareto curves for multi-objective
  optimization problems}.  For example, consider the knapsack problem,
where the input consists of $n$ items with values and sizes.  One
subset of the items dominates another if it has both a larger overall
value and a smaller overall size, and the Pareto curve is defined as
the set of undominated solutions.  Pareto curves matter for
algorithm design because many algorithms for multi-objective
optimization problems (like the Nemhauser-Ullmann
knapsack algorithm) run in time polynomial in the size of the Pareto
curve.  For many problems, the Pareto curve has exponential size in
the worst case but expected polynomial size in a smoothed analysis model.
This chapter also presents a satisfyingly strong connection between
smoothed polynomial complexity and worst-case pseudopolynomial
complexity for linear binary optimization problems.

\subsection{Applications in Machine Learning and Statistics}

Part~V of the book gives a number of examples of how the paradigms in
Parts~I--IV have been applied to problems in {\em machine
  learning and statistics}.

Chapter~16, by Balcan and Haghtalab, considers one of the most basic
problems in supervised learning, that of {\em learning an unknown
halfspace}.  This problem is relatively easy in the noiseless case but
becomes notoriously difficult in the worst case in the presence of
adversarial noise.  This chapter surveys a number of positive
statistical and computational results for the problem under additional
assumptions on the data-generating distribution.  One type of
assumption imposes structure, such as log-concavity, on the marginal
distribution over data points (i.e., ignoring their labels).  A second
type restricts the power of the adversary who introduces the noise,
for example by allowing the adversary to mislabel a point only with a
probability that is bounded away from~$1/2$.

Chapter~17, by Diakonikolas and Kane, provides an overview of recent
progress in {\em robust high-dimensional statistics}, where the goal
is to design learning algorithms that have provable guarantees even
when a small constant fraction of the data points has been adversarially
corrupted.  For example, consider the problem of estimating the
mean~$\mu$ of an unknown one-dimensional Gaussian distribution~$\mathcal{N}(\mu,\sigma^2)$, where the input
consists of $(1-\eps)n$ samples from the distribution and $\eps n$
additional points defined by an adversary.  
The empirical mean of the data points is a good estimator of the true mean
when there is no adversary, but adversarial outliers can distort the
empirical mean arbitrarily.  The median of the input points, however,
remains a good estimator of the true mean even with a small fraction
of corrupted data points.  What about in more than one dimension?
Among other results, this chapter describes a robust and
efficiently computable estimator for learning the mean of a
high-dimensional Gaussian distribution.

Chapter~18, by Dasgupta and Kpotufe, investigates the twin topics of
{\em nearest neighbor search and classification}.  The former is
algorithmic, and the goal is to design a data structure that enables
fast nearest neighbor queries.  The latter is statistical, and the
goal is to understand the amount of data required before the nearest
neighbor classifier enjoys provable accuracy guarantees.  In both
cases, novel parameterizations are the key to narrowing the gap
between worst-case analysis and empirically observed performance---for
search, a parameterization of the data set; for classification, of
the allowable target functions.

Chapter~19, by Vijayaraghavan, is about computing a {\em low-rank
  tensor decomposition}.  For example, given an $m \times n \times p$
3-tensor with entries $\{ T_{i,j,k} \}$, the goal is to express~$T$ as
a linear combination of the minimum-possible number of rank-one
tensors (where a rank-one tensor has entries of the form
$\{ u_i \cdot v_j \cdot w_k \}$ for some vectors $u \in \RR^m$,
$v \in \RR^n$, and $w \in \RR^p$).  Efficient algorithms for this
problem are an increasingly important tool in the design of learning
algorithms; see also Chapters~20 and~21.  This problem is $NP$-hard in
general.
Jennrich's algorithm solves in polynomial time the special case of the
problem in which the three sets of vectors in the low-rank
decomposition (the $u$'s, the $v$'s, and the $w$'s) are linearly
independent.  This result does not address the overcomplete regime,
meaning tensors that have rank larger than dimension.  (Unlike
matrices, the rank of a tensor can be much larger than its smallest
dimension.)  For this regime, the chapter shows that a generalization
of Jennrich's algorithm has smoothed polynomial complexity.

Chapter~20, by Ge and Moitra, concerns {\em topic modeling}, which is
a basic problem in unsupervised learning.  The goal here is to process
a large unlabeled corpus of documents and produce a list of meaningful
topics and an assignment of each document to a mixture of topics.  One
approach to the problem is to reduce it to nonnegative matrix
factorization (NMF)---the analog of a singular value decomposition of
a matrix, with the additional constraint that both matrix factors are
nonnegative.  The NMF problem is hard in general, but this chapter
proposes a condition on inputs, which is reasonable in a topic modeling
context, under which the problem 
can be solved quickly in both theory and practice.  
The key assumption is that each topic has at
least one ``anchor word,'' the presence of which strongly indicates
that the document is at least partly about that topic.

Chapter~21, by Ma, studies the computational mystery outlined in
Section~\ref{ss:ml}: Why are {\em local methods} like stochastic gradient
descent so effective in solving the {\em nonconvex optimization} problems
that arise in supervised learning, such as computing the
loss-minimizing parameters for a given neural network architecture?
This chapter surveys the quickly evolving state-of-the-art on this
topic, including a number of different restrictions on problems under
which local methods have provable guarantees.  For example, some
natural problems have a nonconvex objective function that
satisfies the ``strict saddle condition,'' which asserts that at
every saddle point (i.e., a point with zero gradient that is neither a
minimum nor a maximum) there is a direction with strictly negative
curvature.  Under this condition, variants of gradient descent
provably converge to a local minimum (and, for some problems, a global
minimum).

Chapter~22, by Hardt, tackles the statistical mystery discussed in
Section~\ref{ss:ml}: {\em Why do overparameterized models} like deep
neural networks, which have many more parameters than training data
points, so often {\em generalize} well in practice?  While the jury is
still out, this chapter surveys several of the leading explanations
for this phenomenon, ranging from properties of optimization
algorithms like stochastic gradient descent (including algorithmic
stability and implicit regularization) to properties of data sets
(such as margin-based guarantees).

Chapter 23, by G.\ Valiant and P.\ Valiant, presents two {\em instance
  optimality} results for {\em distribution testing and learning}.
The chapter first considers the problem of learning a discretely
supported distribution from independent samples, and describes an
algorithm that learns the distribution nearly as accurately as would
an optimal algorithm with advance knowledge of the true multiset of
(unlabeled) probabilities of the distribution.  This algorithm is
instance optimal in the sense that, whatever the structure of the
distribution, the learning algorithm will perform almost as well as an
algorithm specifically tailored for that structure.  The chapter then
explores the problem of identity testing: given the description of a
reference probability distribution, $\textbf{p}$, supported on a
countable set, and sample access to an unknown distribution,
$\textbf{q}$, the goal is to distinguish whether
$\textbf{p} = \textbf{q}$ versus the case that $\textbf{p}$ and
$\textbf{q}$ have total variation distance at least $\eps$.  This
chapter presents a testing algorithm that has optimal sample
complexity simultaneously for every distribution~$\textbf{p}$ and
$\eps$, up to constant factors.

\subsection{Further Applications}

The final part of the book, Part~VI, gathers a number of
additional applications of the ideas and techniques introduced in
Parts~I--III.

Chapter~24, by Karlin and Koutsoupias, surveys alternatives to
worst-case analysis in the {\em competitive analysis of online
  algorithms}.  There is a long tradition in online algorithms of
exploring alternative analysis frameworks, and accordingly this
chapter connects to many of the themes of
Parts~I--III.\footnote{Indeed, the title of this book is a riff on
  that of a paper in the competitive analysis of online algorithms
  \citep{KP00}.}  For example, the chapter includes results on
deterministic models of data (e.g., the access graph model for
restricting the allowable page request sequences) and semi-random
models (e.g., the diffuse adversary model to blend worst- and
average-case analysis).

Chapter~25, by Ganesh and Vardi, explores the mysteries posed by the
empirical performance of {\em Boolean satisfiability (SAT) solvers}.
Solvers based on backtracking algorithms such as the
Davis-Putnam-Logemann-Loveland (DPLL) algorithm frequently solve SAT
instances with millions of variables and clauses in a reasonable
amount of time.  This chapter provides an introduction to
conflict-driven clause-learning (CDCL) solvers and their connections
to proof systems, followed by a high-level overview of the
state-of-the-art parameterizations of SAT formulas, including
input-based parameters (such as parameters derived from the
variable-incidence graph of an instance) and output-based parameters
(such as the proof complexity in the proof system associated with CDCL
solvers).

Chapter~26, by Chung, Mitzenmacher, and Vadhan, uses ideas from
pseudorandomness to explain {\em why simple hash functions work} so
well in practice.  Well-designed hash functions are practical proxies
for random functions---simple enough to be efficiently implementable,
but complex enough to ``look random.''  In the theoretical analysis of
hash functions and their applications, one generally assumes that a
hash function is chosen at random from a restricted family, such as a
set of universal or $k$-wise independent functions for small~$k$.  For
some statistics, such as the expected number of collisions under a
random hash function, small families of hash functions provably
perform as well as completely random functions.  For others, such as
the expected insertion time in a hash table with linear probing, simple
hash functions are provably worse than random functions (for
worst-case data).  The running theme of this chapter is that a little
randomness in the data, in the form of a lower bound on the entropy
of the (otherwise adversarial) data-generating
distribution, compensates for any missing randomness in a universal
family of hash functions.

Chapter~27, by Talgam-Cohen, presents an application of the beyond
worst-case viewpoint in algorithmic game theory, to {\em
  prior-independent auctions}.  For example, consider the problem of
designing a single-item auction, which solicits bids from bidders and
then decides which bidder (if any) wins the item and what everybody
pays.  The traditional approach in economics to designing
revenue-maximizing auctions is average-case, meaning that the setup
includes a commonly known distribution over each bidder's willingness
to pay for the item.  An auction designer can then implement an
auction that maximizes the expected revenue with respect to the
assumed distributions (e.g., by setting a distribution-dependent
reserve price).  As with many average-case frameworks, this approach
can lead to impractical solutions that are overly tailored to the
assumed distributions.  A semi-random variant of the model 
allows an adversary to pick its favorite distribution out of a rich
class, from which nature chooses a random sample for each bidder.  
This chapter presents prior-independent auctions, both with and
without a type of resource augmentation, that achieve near-optimal
expected revenue simultaneously across all distributions in the
class.

Chapter~28, by Roughgarden and Seshadhri, takes a beyond worst-case
approach to the {\em analysis of social networks}.  Most
research in social network analysis revolves around a collection of
competing generative models---probability distributions over graphs
designed to replicate the most common features observed in such
networks.  The results in this chapter dispense with generative models
and instead provide algorithmic or structural guarantees under
deterministic combinatorial restrictions on a graph---that is, for
restricted classes of graphs.  The restrictions are motivated by the
most uncontroversial properties of social and information networks,
such as heavy-tailed degree distributions and strong triadic closure
properties.  Results for these graph classes effectively apply
to all ``plausible'' generative models of social
networks.

Chapter~29, by Balcan, reports on the emerging area of {\em
  data-driven algorithm design}.  The idea here is to
model the problem of selecting the best-in-class algorithm for a given
application domain as an offline or online learning problem, in the
spirit of the aforementioned work on self-improving algorithms.  For
example, in the offline version of the problem, there is an unknown
distribution $D$ over inputs, a class~$C$ of allowable algorithms, and
the goal is to identify from samples the algorithm in~$C$ with the
best expected performance with respect to $D$.  The distribution~$D$
captures the details of the application domain, the samples correspond
to benchmark instances representative of the domain, and the
restriction to the class~$C$ is a concession to the reality that it is
often more practical to be an educated client of already-implemented
algorithms than to design a new algorithm from scratch.  For many
computational problems and algorithm classes~$C$, it is possible to
learn an (almost) best-in-class algorithm from a modest number of
representative instances.

Chapter~30, by Mitzenmacher and Vassilvitskii, is an introduction to
{\em algorithms with predictions}.  For example, in the online paging
problem (Section~\ref{ss:paging}), the LRU policy makes predictions
about future page requests based on the recent past.  If its
predictions were perfect, the algorithm would be optimal.  What if a
good but imperfect predictor is available, such as one computed by a
machine learning algorithm using past data?  An ideal solution would
be a generic online algorithm that, given a predictor as a ``black
box'': (i) is optimal when predictions are perfect; (ii) has
gracefully degrading performance as the predictor error increases; and
(iii) with an arbitrarily poor predictor, defaults to the optimal
worst-case guarantee.  This chapter investigates the extent to which
properties~(i)--(iii) can be achieved by predictor-augmented
data structures and algorithms for several different problems.

\section{Notes}

This chapter is based in part on \citet{cacm}.

The simplex method (Section~\ref{ss:lp}) is described, for example, in
\citet{D63}; \citet{K79} proved that the ellipsoid method solves
linear programming problems in polynomial time;
and the first polynomial-time interior-point method was developed by
\citet{K84}.
Lloyd's algorithm for $k$-means (Section~\ref{ss:clustering})
appears in \citet{lloyd}.
The phrase ``clustering is hard only when it  doesn't matter'' (Section~\ref{ss:clustering})
is credited to Naftali Tishby  by~\citet{DLS12}.
The competitive analysis of online algorithms
(Section~\ref{ss:paging}) was pioneered 
by \citet{ST85}.  B\'el\'ady's algorithm (Section~\ref{ss:paging})
appears in \citet{B67}.
The working set model in Section~\ref{ss:workingset} was formulated by
\citet{D68}.  Theorem~\ref{t:afg} is due to \citet{AFG02}, as is
Exercise~\ref{ex:fifo}.  Exercise~\ref{ex:param1} is folklore.
The result in Exercise~\ref{ex:perceptron} is due to \citet{B62} and
\cite{N62}.

\section*{Acknowledgments}

I thank J\'er\'emy Barbay, Daniel Kane, and Salil Vadhan
for helpful comments on a preliminary draft of this chapter.

\section*{Exercises}

\begin{enumerate}

\item \label{ex:paging}
Prove that for every deterministic cache replacement policy and cache
size~$k$, there is an adversarial page request sequence such that the
policy faults on every request, and such that an optimal clairvoyant
policy would fault on at most a~$1/k$ fraction of the requests.

\vspace{.25\baselineskip}
\noindent
[Hint: use only $k+1$ distinct pages, and the fact that the optimal
policy always evicts the page that will be requested furthest in the
future.]

\item \label{ex:concave}
Let~$f:\NN \rightarrow \NN$ be a function of the type described in
Section~\ref{s:afg}, with~$f(n)$ denoting the maximum allowable number
of distinct page requests in any window of length~$n$.
\begin{exlist}

\item Prove that there is a nondecreasing function~$f':\NN
  \rightarrow \NN$ with $f'(1)=1$
  and $f'(n+1) \in \{ f'(n),f'(n+1) \}$ for all~$n$ such that a
  page request sequence conforms to~$f'$ if and only if it conforms to
  $f$.

\item Prove that parts~(a) and~(b) of Theorem~\ref{t:afg} hold
  trivially if~$f'(2) = 1$.

\end{exlist}

\item \label{ex:conform}
Prove that the page request sequence constructed in the proof of
Theorem~\ref{t:afg}(a) conforms to the given approximately concave
function~$f$.

\begin{figure}
\begin{center}
\begin{tabular}{c|c|c|c|c|c|c|c|c}
$f(n)$ & 1 & 2 & 3 & 3 & 4 & 4 & 5 & 5  \\ \hline
   $n$ & 1 & 2 & 3 & 4 & 5 & 6 & 7 & $\cdots$ \\
\end{tabular}
\caption{Function used to construct a bad page sequence for FIFO (Exercise~\ref{exer:afg}).}\label{f:ctrex}
\end{center}
\end{figure}

\item \label{exer:afg}
Prove Theorem~\ref{t:afg}(c).

\vspace{.25\baselineskip}
\noindent [Hint: Many different choices of $f$ and $k$ work.  For
example, take $k=4$, a set $\{0,1,2,3,4\}$ of 5 pages, the function
$f$ shown in Figure~\ref{f:ctrex}, and a page request sequence
consisting of an arbitrarily large number of identical blocks of the
eight page requests 10203040.]

\item \label{ex:fifo}
Prove the following analog of Theorem~\ref{t:afg}(b) for the FIFO
replacement policy: for every $k \ge 2$ and approximately concave
function $f$ with $f(1)=1$, $f(2) = 2$, and $f(n+1) \in \{ f(n),
f(n+1) \}$ for all~$n \ge 2$, the page 
fault rate of the FIFO policy on every request sequence that conforms
to $f$ is at most
\begin{equation}\label{eq:fifo}
\frac{k}{f^{-1}(k+1)-1}.
\end{equation}

\vspace{.25\baselineskip}
\noindent
[Hint: make minor modifications to the proof of Theorem~\ref{t:afg}(b).
The expression in~\eqref{eq:fifo} suggests defining phases
  such that (i) the FIFO policy makes at most $k$ faults per phase;
  and (ii) a phase plus one additional request 
  comprises requests for at least $k+1$ distinct pages.]

\item \label{ex:param1}
An instance of the knapsack problem consists of
$n$ items with
nonnegative values $v_1,\ldots,v_n$ and sizes $s_1,\ldots,s_n$,
and a knapsack capacity $C$.  The goal is to compute a
subset $S \sse \{1,2,\ldots,n\}$ of items that fits in the knapsack
(i.e., with $\sum_{i \in S} s_i \le C$) and, subject to this, has the
maximum total value $\sum_{i \in S} v_i$.

One simple greedy algorithm for the problem
reindexes the items in nonincreasing order of
density~$\tfrac{v_i}{s_i}$ and then returns the largest prefix $\{1,2,\ldots,j\}$ of
items that fits in the knapsack (i.e., with $\sum_{i=1}^j s_i \le C$).
Parameterize a knapsack instance by the ratio~$\alpha$ of the largest
size of an item and the knapsack capacity, and
prove a parameterized guarantee for the greedy algorithm: the total value of
its solution is at least $1-\alpha$ times that of an optimal solution.

\vspace{.25\baselineskip}

\begin{figure}[t]
\hrule\medskip
\begin{flushleft}
\textbf{Input}: $n$ unit vectors $\bfx_1,\ldots,\bfx_n \in \RR^d$ with labels
$b_1,\ldots,b_n \in \{-1,+1\}$.

\begin{enumerate}

\item Initialize $t$ to~1 and $\bfw_1$ to the all-zero vector.

\item While there is a point~$\bfx_i$ such that
$\sgn(\bfw_t \cdot \bfx_i) \neq b_i$,
set~$\bfw_{t+1} = \bfw_{t} + b_i\bfx_i$ and
increment~$t$.\tablefootnote{Intuitively, this update step forces
the next vector to be ``more correct'' on~$\bfx_i$, by increasing
$\bfw \cdot \bfx_i$ by $b_i(\bfx_i \cdot \bfx_i) = b_i$.}

\end{enumerate}
\caption{\textsf{The Perceptron Algorithm.}\label{fig:perceptron}}
\end{flushleft}
\medskip
\hrule
\end{figure}

\item \label{ex:perceptron}
The {\em perceptron algorithm} is one of the most classical machine
learning algorithms (Figure~\ref{fig:perceptron}).
The input to the algorithm is $n$ points in $\RR^d$,
with a label $b_i \in \{-1,+1\}$ for each point $\bfx_i$.  The goal is to
compute a {\em separating hyperplane}: a hyperplane with all of the
positively-labeled points on one side, and all of the
negatively labeled points on the other.  
Assume that there exists a separating hyperplane, and moreover that
some such hyperplane passes
through the origin.\footnote{The second assumption is without loss of
  generality, as it can be enforced by adding an extra ``dummy
  coordinate'' to the data points.}
We are then free to scale each data point $\bfx_i$ so that $\|\bfx_i\|_2=1$---this
does not change which side of a hyperplane $\bfx_i$ is on.

Parameterize the input by its {\em margin} $\mu$, defined as
\[
\mu = \max_{\bfw \,:\, \n{\bfw}=1} \min_{i=1}^n |\bfw \cdot \bfx_i|,
\]
where $\bfw$ ranges over the unit normal vectors of all separating hyperplanes.
Let~$\bfw^*$ attain the maximum.
Geometrically, the parameter~$\mu$ is the smallest cosine of an angle
defined by a point $\bfx_i$ and the 
normal vector~$\bfw^*$.

\begin{exlist}

\item
Prove that the squared norm of $\bfw_t$ grows slowly with the
number of iterations~$t$:
$
\n{\bfw_{t+1}}^2 \le \n{\bfw_t}^2 + 1
$
for every~$t \ge 1$.

\item
Prove that the projection of $\bfw_t$ 
onto~$\bfw^*$ grows significantly with every iteration:
$
\bfw_{t+1} \cdot \bfw^* \ge \bfw_t \cdot \bfw^* + \mu
$
for every~$t \ge 1$.

\item
Conclude that the iteration count~$t$ never exceeds $1/\mu^2$.

\end{exlist}

\end{enumerate}

\end{document}